\documentstyle[prl,aps]{revtex}
\begin{document}
\draft
\voffset=.5in
\twocolumn[\hsize\textwidth\columnwidth\hsize\csname
@twocolumnfalse\endcsname

\title
{Cutoff dependence and Lorentz invariance of the Casimir effect}

\author {C. R. Hagen\cite{Hagen}}

\address
{Department of Physics and Astronomy\\
University of Rochester\\
Rochester, N.Y. 14627}

\maketitle

\begin{abstract}
The Casimir force between two conducting planes is considered in both the
electromagnetic and scalar field cases.  This is done by the usual
summation over energy eigenmodes of the system as well as by a calculation of
the stress tensor in the region between the planes.  The latter case requires
that careful attention be given to singular operator products, an issue which
is accommodated here by invoking the point separation method in conjunction
with a scalar cutoff.  This is shown to yield cutoff dependent and divergent
contributions to the Casimir pressure which are dependent on the separation
parameters, but entirely consistent with Lorentz covariance.  Averaging over
the point splitting parameters allows finite results to be obtained, but fails
to yield a unique Casimir force.
\end{abstract}

\pacs {11.10Kk, 12.20.Ds}

\vskip2pc]
\bigskip

     In 1948 Casimir [1] first predicted that two infinite parallel plates in
vacuum would attract each other.  This force has its origin in the vacuum zero
point energy and has long been an object of intense scrutiny from both the
experimental and theoretical points of view.  With regard to the former it is
generally well known that the relatively large experimental uncertainties of
early experiments [2] allowed one only to conclude that the reality of the
effect could not be excluded.  While more recent experimental work [3] has
produced data which is considerably more impressive, it is far from clear that
the Casimir effect has been unambiguously confirmed.  This has to do with the
fact that these recent experiments are based on a different geometry from that
of Casimir.  Comparison of data with the Casimir prediction has been based
upon the invocation of the \lq\lq proximity force theorem", an approach which 
requires the existence of a well defined potential function.  Since the 
calculation of the Casimir effect requires a summation over all the allowed 
modes of the radiation field, such a function is not available and the 
relevance of the experiments [3] to the Casimir effect remains uncertain.

The present work, however, is concerned solely with theoretical
issues which relate to the Casimir effect.   Because of this it is well to
remark at the outset that experimental and theoretical aspects can (and
probably should) be considered independent of each other, despite the fact that
the {\it ultimate} goal must, of course, be to reconcile them.  Thus it can be
imagined that fresh experiments might well lead to a confirmation of the
Casimir prediction even though (as shown in this work) there remain
significant unresolved theoretical issues.  In that case one might well take
the position that some new theoretical insight (which may or may not have to do
with zero point energy) would be required.

The most direct calculation of the Casimir force relies on mode summation.
Indeed it is precisely this calculational technique which is usually invoked
when the goal is to derive the predicted results within a physically plausible
framework.  In this approach one calculates the summation of ${1\over 2}\omega$
[4] where the sum is extended over all allowed modes of the system.  This is a
highly divergent result, going as the fourth power of a cutoff frequency in
three dimensions and a regularization method must be prescribed.  It is known
[5] in the case of parallel plate geometry that the final result is independent
of the details of regularization so long as the relevant cutoff function
depends only on the mode frequency.  However, this partial result certainly
cannot be construed as a general proof of the cutoff independence of the
Casimir force.  Since efforts to provide such a proof have thus far not been
successful, it is natural to attempt to determine whether there exists the
possibility of finding alternative regularizations which violate the condition
of cutoff independence.

To this end one can consider the case of the electromagnetic field in three
dimensions with conducting plates at $z=0,a$.  The ground state energy is
formally given by $E=\sum \omega$
where the expected factor of ${1\over 2}$ has been omitted as a consequence of
the fact that the photon field has two independent polarizations.  In order to
lend meaning to the above expression a regularization is, of course, required.
Thus one for $\cal E$ (the energy per unit area of the plates) one can write
\begin{equation}
    {\cal E} =- \sum^{\prime} \int {d^2k\over (2\pi)^2}
   \exp [ \lambda\varepsilon{n\pi\over a} ]
    {\partial\over\partial\varepsilon}
   \exp [ -\varepsilon\omega_k ]
\end{equation}
where $\omega_k^2={\bf k}^2+({n\pi\over a})^2$ prescribes the allowed modes of
the system and cutoff parameters $\varepsilon$ and $\lambda$ have been
introduced.  A prime has been placed on the summation to indicate
that for $n=0$ there is only a single polarization thus requiring that the
$n=0$
term be reduced by one-half.  The Casimir energy is to be evaluated in the
limit $\varepsilon\to 0$.  The usual approach in which one takes
$\lambda=0$ from the outset marginally expedites the calculation, but clearly
foregoes any possibility of examining the effects of more general cutoff
schemes.

The integral over the transverse coordinates and the sum over $n$ are readily
carried out to yield
$${\cal E} (\varepsilon,\lambda) = {\partial^2\over \partial \varepsilon^2}
{1\over 4\pi\varepsilon}
\coth [ (\varepsilon -\lambda\varepsilon'){\pi\over 2a} ] |_{\varepsilon=
\varepsilon'}.$$
Upon evaluating the derivatives and expanding around $\varepsilon =0$ this
yields
\begin{eqnarray*}
    {\cal E} & = & -{\lambda\over 12a\varepsilon^2}+a{(2-\lambda)(1-\lambda)+
{2\over 3}\over \pi^2\varepsilon^4(1-\lambda)^3}+{\pi^2\over 720a^3}
\lambda (\lambda^2-1) \nonumber \\
& - & (1-\lambda){\pi^2\over 720a^3}+{a\over 3\pi^2}
{1\over \varepsilon^4}{1\over (1-\lambda)^3}.
\end{eqnarray*}
Since one is ultimately interested into the case in which the conducting
plate at $z=a$ is an interior wall of the system, one obtains the relevant
Casimir energy $\cal E$ by adding the corresponding result for $a\to L-a$ with
$L\gg a$.  Upon discarding irrelevant $a$ independent terms this yields
\begin{equation}
\overline{\cal E}= -(1-\lambda){\pi^2\over 720a^3}-{\lambda\over 12a\varepsilon
^2}
+{\pi^2\over 720a^3}\lambda(\lambda^2-1).
\end{equation}
One sees immediately from this expression that the Casimir energy reduces to
the commonly quoted result $\overline {\cal E} = -{\pi^4\over 720a^3}$ in
the
limit $\lambda=0$.  However, in the case of nonzero $\lambda$, the result found
here is divergent and explicitly cutoff dependent, an observation which applies
equally to its gradient (i.e., the negative of the Casimir pressure).

The calculation for the case of a scalar field required to satisfy Dirichlet
boundary conditions at $z=0,a$ proceeds identically, but with the modification
that there is one less mode for each value of $n$.  Thus the result is simply
one-half that given by Eq.(1) provided that the prime on the summation is now
taken to require the total deletion of the $n=0$ term.  Since the latter in any
event does not contribute to the Casimir energy, one finds without further
calculation that the Casimir enegy in the scalar case is simply one-half that
given by Eq.(2).  This simple correspondence between the electromagnetic and
scalar cases will be found not to persist when one examines the form of
the stress tensor in the region between the plates for these two fields.

This result demonstrates that the a simple and straightforward modification of
the usual regularization method results in a total and unexpected change in the
predicted value for the force on the plates.  It is not farfetched in the light
of this result to anticipate that this will be generally true of Casimir
effect calculations.  These have, for very practical reasons of viability,
been largely limited to exponential cutoff functions.  Yet it has been found
that even within this narrow class there exists the possibility of
demonstrating cutoff dependence.  On the other hand it has been known in more
recent years that there exists the possibility of approaching the parallel
plate Casimir effect by means of the stress tensor.  This raises the additional
issue of whether Lorentz invariance in $2+1$ dimensions (i.e., $O(2,1)$
invariance) could provide additional information which bears upon this issue.
Specifically, could it serve to place limits on the class of admissible cutoff
functions which need to be considered?  The remainder of this work will seek
to resolve precisely this issue.

The Casimir effect for parallel plate geometry was formulated in terms of the
vacuum expectation of the stress tensor $T^{\mu\nu}$ by Brown and Maclay [6]
using an image method.  Basically this approach strives to satisfy the
appropriate boundary conditions at the plates by showing that the photon
propagator between the two plates can be expressed as an infinite sum over the
usual (i.e., $-\infty<z<\infty)$ propagators in which the successive terms
in the
series have their $z$-coordinates displaced in accord with the standard image
method.  Using the fact that the stress tensor for the electromagnetic field
has the traceless form
\begin{equation}
T^{\mu\nu}(x)= F^{\mu\alpha} F^{\nu}_{\;\alpha}-{1\over4}
g^{\mu\nu}F^{\alpha\beta}F_{\alpha\beta}
\end{equation}
where $F^{\mu\nu}(x)=\partial^{\mu}A^{\nu}(x)-\partial^{\nu}A^{\mu}(x),$
it follows that upon taking appropriate derivatives with respect to the
propagator arguments $x$ and $x'$ and invoking the limit $x \rightarrow x'$ a
formal expression can be obtained for the vacuum expectation value of the
stress tensor.  Following this procedure it was found [6] that the vacuum
expectation of the stress tensor has the form
\begin{equation}
\langle 0|T^{\mu\nu}(x)|0\rangle =({1\over 4}g^{\mu\nu}-\hat{z}^{\mu}
\hat{z}^{\nu})({1\over 2\pi^2a^4})\sum_{n=1}^{\infty}n^{-4}
\end{equation}
where $\hat{z}^{\mu}$ is the unit vector (0,0,1,0) in the $z$-direction normal
to the conducting planes.

Despite the simplicity and elegance of the approach of [6], there is some
reason to suggest the necessity of a reexamination of this calculation.  One
notes first of all that the obviously singular $n=0$ term in (4) has been
deleted.  This has been done by invoking the argument that such a term
represents the large $a$ contribution which should, of course, be subtracted
since the physically observable quantity is the difference between the finite
and infinite $a$ results.  This term is biquadratically divergent and 
nonleading terms in the calculation of such a quantity can be notoriously
dependent on the specifics of the regularization procedure.  There is the
additional concern that the deletion of the $n=0$ term has the effect of
invalidating the boundary conditions on the conducting plates.   This is a
simple consequence of the fact that {\it all} the terms in the series are
required to ensure that the boundary conditions are rigorously satisfied.
A final point has to do with the fact that [6] assumes from the outset the
absence of divergences in (4), a result which, if indeed true, should emerge
as a consequence of the calculation itself rather than as input.

To avoid such difficulties this work approaches the problem through the
technique of mode summation.  This has the advantage that each contribution to
the stress tensor unambiguously satisfies the boundary conditions on the
plates, thereby reducing the problem to one of how the regularization of the
infinite sum should be carried out.  This approach has previously been used by
DeWitt [7] and the author [8].  The former work uses only the usual $\omega$
dependent regularization to derive Eq.(4) thereby leaving open the issue of
more general cutoff approaches.  In [8] an $O(2,1)$ vector cutoff $\sigma^\mu$
was used which indeed led to a cutoff dependence, but left unresolved the issue
of whether the dependence of the results on this external vector could be
reconciled with Lorentz invariance.  By using here a definition of the stress
tensor based on standard point splitting methods of field theory the covariance
is issue readily seen to be resolved.

To carry out this program one begins with an expansion of the Green's function
in terms of orthogonal functions [9].  In particular the free field propagator
in the radiation gauge can be written as

\begin{eqnarray}
G^{ij}({\bf x-x'},z,z',t-t')
= \sum_{n\lambda}\int {d{\bf k} d\omega \over (2\pi)^3}
e^{-i\omega (t-t')} \nonumber \\
\times
{ A_{n\lambda}^i({\bf k},z) A_{n\lambda}^{j*}
({\bf k},z')\over k^2-{\omega}^2+(n\pi/a)^2-i\epsilon}
e^{i{\bf k}\cdot ({\bf x-x'})}
\end{eqnarray}
where $\lambda =1,2$ refers to the polarization, and spatial coordinates
orthogonal to the $z$-direction are denoted by a boldface notation.  The
eigenfunctions $A_{n\lambda}^i({\bf k},z)$ are given explicitly by
\begin{equation}
A_{n1}^i({\bf k},z)={\overline{k}_i\over |{\bf k}|}({2\over a})^{1\over 2}
\sin(n\pi z/a)
\end {equation}
and
\begin{equation}
A_{n2}^i({\bf k},z)={1\over |{\bf k}|\omega_k}
\left(\hat{\bf z}^i{\omega}_k^2+
\hat{{\bf z}}\cdot{\bf \nabla}{\nabla}^i\right)
({2\over a})^{1\over 2}\cos(n\pi z/a)
\end{equation}
where $\overline{k}_i \equiv \epsilon^{ij}k_j$ with $\epsilon^{ij}$ being the
usual alternating symbol.  In the $n=0$ case normalization requires that the
rhs of (7) must be multiplied by a factor of $2^{-{1\over 2}}$.  It is
important to note that each eigenfunction $A_{n\lambda}^i({\bf k},z)$ satisfies
the boundary conditions $\hat{{\bf z}}\times {\bf E}=\hat{{\bf z}}\cdot{\bf B}=
0$ at $z=0,a$.  This means that it is possible to introduce a regularization
such that contributions from large values of $|{\bf k}|$ and/or $n$ are
reduced
without destroying the validity of the boundary conditions.  This stands in
marked contrast with the image method which has no mechanism for the consistent
suppression of the contributions of higher order reflections.

A mathematically well defined vacuum stress is defined by the relation [10]
\begin{eqnarray*}
\langle 0 & | & T^{\mu\nu}(\varepsilon)|0\rangle_\lambda =
(\delta^\mu_\alpha\delta^\nu
_\beta-{1\over 4}g^{\mu\nu}g_{\alpha\beta}) \nonumber \\
& & \Re \langle 0|[F^{\alpha\sigma}
(x+{\varepsilon\over 2})\exp\{\lambda[-P^2\varepsilon^2]^{1\over
2}\}F^\beta_\sigma
(x-{\varepsilon\over2})|0\rangle
\end{eqnarray*}
where $P^2\equiv {\bf P}^2-E^2$ with $E$ and ${\bf P}$ are respectively the
energy and momentum operators associated with the $2+1$ dimensional subspace.
The form of the $\lambda$-dependent term has been chosen to agree with results
obtained earlier for $\cal E$.  By straightforward
 calculation this is reduced to
\begin{eqnarray}
\langle 0 & | & T^{\mu\nu}(\varepsilon)|0\rangle_\lambda  = \Re {-2i\over a}
{\sum_{n=0}^{\infty}}^\prime
\int { d^3k \over (2\pi)^3}
{1\over k^2 + (n\pi/a)^2-i\epsilon} \nonumber \\
& & \times  e^{ik^\mu\varepsilon_\mu}e^{\lambda\varepsilon n\pi/a}
\left[ k^\mu k^\nu +\hat{z}^\mu \hat{z}^\nu (n\pi/a)^2 \right]
\end{eqnarray}
which is manifestly both symmetric and traceless.  It can be more usefully
written as
\begin{eqnarray}
\langle 0 &| &T^{\mu\nu}(\varepsilon)|0\rangle_\lambda =\Re{-2i\over
a}{\sum_{n=0}
^{\infty}}^\prime e^{\lambda\varepsilon n\pi/a} \nonumber \\
& & \times \left( -{\partial\over \partial \varepsilon_\mu}
{\partial\over \partial\varepsilon_\nu}
+\hat{z}^\mu \hat{z}^\nu {\partial^2\over \partial\varepsilon^{\alpha}
\partial\varepsilon_{\alpha}} \right) \Delta^{n\pi/a}(\varepsilon)
\end{eqnarray}
where $\Delta^{n\pi/a}(x)$ is the (2+1) dimensional function
$$\Delta^{n\pi/a}(x)=\int{d^3k\over (2\pi)^3}e^{ikx}
{1\over k^2 +(n\pi/a)^2-i\epsilon}$$
for a particle of mass $n\pi/a$.  Since this is an $O(2,1)$ scalar,
$\Delta^{n\pi/a}(\varepsilon)$ is a function of only the invariant
${\varepsilon}^2$ and has the explicit form
$$\Delta^{n\pi/a} (\varepsilon) = {i\over 4\pi\varepsilon}
e^{-\varepsilon n\pi/a}.$$

The summation over $n$ can now readily be performed to obtain
$$
\langle 0|T^{\mu\nu}(\varepsilon)|0\rangle_\lambda=
(-{\partial\over \partial\varepsilon_\mu}
{\partial\over \partial\varepsilon_\nu}+\hat{z}^{\mu}
\hat{z}^{\nu}{\partial^2\over \partial\varepsilon^{\alpha}\partial\varepsilon_
{\alpha}})F(\varepsilon,\varepsilon',\lambda)|_{\varepsilon'=\varepsilon}$$
where
$$F(\varepsilon,\varepsilon',\lambda) = {1\over 4\pi a\varepsilon}
\coth [(\varepsilon-\lambda\varepsilon'){\pi\over 2a}].$$
In the limit of small $\varepsilon$ the latter reduces to
$$F(\varepsilon,\varepsilon',\lambda)\rightarrow
\left[{1\over 2\pi^2}{1\over\varepsilon}{1\over
\varepsilon-\lambda\varepsilon'}-
{\lambda\varepsilon'\over
24a^2\varepsilon}-{(\varepsilon-\lambda\varepsilon')^3\pi^2\over
1440\varepsilon a^4}
\right].$$
Upon performing the derivatives and subtracting the $a\to\infty$ result one
finds
\begin{eqnarray}
\langle 0 & | & \overline{T}^{\mu\nu}(\varepsilon)|0\rangle_\lambda =
\left( {1\over 4}g^{\mu\nu}-\hat{z}^{\mu}\hat{z}^{\nu}\right)
( 1-\lambda) {\pi^2\over 180a^4} \nonumber \\
&  &  \left[ g^{\mu\nu} -
3 {\varepsilon^{\mu}\varepsilon^{\nu}\over \varepsilon^2}
- \hat{z}^{\mu}\hat{z}^{\nu} \right]
 \lambda \left\{ - {1\over 24a^2\varepsilon^2}
+ {\pi^2\over 1440a^4}
 { ( \lambda^2-1 ) } \right\}
\end{eqnarray}
where an overbar notation has been used to denote this subtraction.

Since the vacuum stress obtained here is dependent on the components of the
three-vector $\varepsilon^\mu$, it is necessary to demonstrate that there
is no
contradiction of Lorentz covariance implied by this result.  To this end one
merely needs to note that for a $2+1$ dimensional Lorentz transformation
$x^{\mu\prime}=\ell^\mu_\nu x^\nu$ generated by the unitary operator $U$ the
electromagnetic field tensor transforms as
$$UF^{\mu\nu}(x)U^\dagger=\ell^\mu_\alpha\ell^\nu_\beta F^{\alpha\beta}
(\ell^{-1}x).$$
This clearly implies that
$$\langle 0|UT^{\mu\nu}(\varepsilon)U^\dagger|0\rangle_\lambda=\ell^\mu_\alpha
\ell^\nu_\beta
\langle 0|T^{\alpha\beta}(\ell^{-1}\varepsilon)|0\rangle_\lambda.$$
Since the result (10) is in accord with this transformation law, it follows
that there is no implied conflict with Lorentz invariance.  If, however, it is
desired that there should be no explicit reference to the components
$\varepsilon^\mu$ [11], an averaging over all directions can be carried out to
obtain
\begin{equation}
\langle 0|\overline{T}^{\mu\nu}(\varepsilon)|0\rangle_\lambda^{ave} =
({1\over 4}g^{\mu\nu}-\hat{z}^\mu\hat{z}^\nu)(1-\lambda){\pi^2\over 180a^4}.
\end{equation}
While this result has the desired feature of being finite and manifestly
covariant, it demonstrates the principal claim of this paper concerning the
dependence of the Casimir energy upon the particular regularization employed.

The scalar field case shares a number of features with the electromagnetic one,
but somewhat unexpectedly displays the fact that at least for certain
regularizations the vacuum stress can be $z$-dependent.  One starts from the
traceless form of the stress tensor
$$T^{\mu\nu}=\phi^\mu\phi^\nu-{1\over 2}g^{\mu\nu}\phi_\alpha^2+{1\over 6}
(g^{\mu\nu}\partial^2-\partial^\mu\partial^\nu)\phi^2$$
where $\phi^\mu=-\partial^\mu\phi$ and imposes Dirichlet boundary conditions at
$z=0,a$.  In this case (7) becomes
\begin{eqnarray*}
\langle 0 &| &T^{\mu\nu}(\varepsilon)|0\rangle_\lambda =
\Re{-2i\over a}\sum_{n=1}^\infty\int{d^3k\over (2\pi)^3}
{k^2\over k^2 + (n\pi/a)^2-i\epsilon}
\nonumber \\
&  & e^{ik^\mu\varepsilon_\mu} e^{\lambda\varepsilon n\pi/a}
\left[ {2\over 3}
\left({1\over 4}g^{\mu\nu}-\hat{z}^\mu\hat{z}^\nu \right) +  \nonumber
\right.\\
&  & \left. \left( {k^\mu k^\nu\over k^2} - {1\over 3} g^{\mu\nu} +
{1\over 3}\hat{z}^\mu \hat{z}^\nu \right) \sin^2 {n\pi z\over a} \right].
\end{eqnarray*}
The once subtracted form of the vacuum stress is then found to be of the form
[12]
\begin{eqnarray*}
    \langle 0 & | & \overline{T}^{\mu\nu} (\varepsilon)|0\rangle_\lambda =
    ({1\over 4}g^{\mu\nu}- \hat{z}^\mu\hat{z}^\nu)
(1-\lambda){\pi^2\over 360a^4} +  \\
& & {\lambda\over 48}
\left( g^{\mu\nu}-3{\varepsilon^\mu\varepsilon^\nu\over \varepsilon^2} -
\hat{z}^\mu\hat{z}^\nu \right)
\left\{ {1\over a^2\varepsilon^2}
\left( {3\over \sin^2{\pi z\over a}}-1 \right) \right. \\
& & \left. +{\pi^2\over 4a^4}\left( 1-\lambda^2 \right)
\left[ {3-2\sin^2{\pi z\over a}\over \sin^4 {\pi z\over a}}
-{1\over 15} \right] \right\},
\end{eqnarray*}
a result which is in itself problematical in that it implies the existence of
nonintegrable singularities in the expression for the total Casimir energy.
Averaging over directions of $\varepsilon$ yields a $\lambda$ dependent result
precisely half that given by (11) for the electromagnetic case.

In sum, it has been shown here that the Casimir effect for plane geometry quite
generally leads to a regularization dependence of the Casimir force and that
such results follow quite naturally from point splitting techniques commonly
used in field theory.  Similar conclusions have been obtained [8] in the case
of the sphere.  These observations constitute a significant basis on which to
question the physical reality of the Casimir effect as an observable
consequence of vacuum zero point energy.

\acknowledgments

This work is supported in part by the U.S. Department of Energy Grant
No.DE-FG02-91ER40685.


\begin{references}
\bibitem{ } The units employed here are such that $\hbar=c=1$ with the
signature of the metric being (1,1,1,-1).
\bibitem{ } To avoid complications associated with the boundary conditions on
the conducting planes point splitting is carried out only in the $2+1$
dimensional subspace $\varepsilon^\mu\hat{z}_\mu=0$.
\bibitem{ } It must be emphasized, however, that such a requirement represents
only one specific choice and is not something required by the underlying field
theory.
\bibitem{ } In [7] similar $z$-dependent terms which cause the integral of the
vacuum stress to diverge in the scalar case have been found to occur if one
uses the so-called \lq\lq unimproved" form $T^{\mu\nu}=\phi^\mu\phi^\nu
-{1\over 2}g^{\mu\nu}\phi^2_\alpha$.  The non-occurrence of $z$-dependent terms
for the \lq\lq improved" stress tensor found in [7]is thus seen here not to be 
a general characteristic independent of the regularization technique.
\end{references}
\end{document}